\documentclass[12pt,preprint]{aastex}
\usepackage{natbib}
\newcommand{\Msun}{M{$_{\odot}$}}

\newcommand{\kms}{km~s{$^{-1}$}}

\newcommand{\ha}{\mbox{H$\alpha$}}
\newcommand{\sii}{[S{\sc ii}]}
\newcommand{\oii}{[O{\sc ii}]}
\newcommand{\mh}{H$_2$}
\newcommand{\Ks}{K$_S$}
\newcommand{\lkha}{LkH$\alpha$~}
\newcommand{\um}{$\mu$m}

\slugcomment{}

\shorttitle{L988}
\shortauthors{Walawender et al.}

\begin{document}

\title{Optical and Near-Infrared Shocks in the L988 Cloud Complex}

\author{J.\ Walawender}
\affil{Subaru Telescope, National  Astronomical Observatory of Japan, Hilo, HI 96720}
\email{joshw@naoj.org}
\author{B.\ Reipurth}
\affil{Institute for Astronomy, University of Hawaii at Manoa, Hilo, HI 96720}
\and
\author{J.\ Bally}
\affil{Center for Astrophysics and Space Astronomy, University of Colorado, Boulder, CO 80309}

\begin{abstract}
We have searched the Lynds~988 dark cloud complex for optical (\ha{} and \sii{}) and near-IR (\mh{} 2.12 $\mu$m) shocks from protostellar outflows.  We find 20 new Herbig-Haro objects and 6 new \mh{} shocks (MHO objects), 3 of which are cross detections.  Using the morphology in the optical and near-IR, we connect several of these shocks into at least 5 distinct outflow systems and identify their source protostars from catalogs of infrared sources.

Two outflows in the cloud, from IRAS~21014+5001 and IRAS~21007+4951, are in excess of 1~pc in length.  The IRAS~21007+4951 outflow has carved a large cavity in the cloud through which background stars can be seen.  Also, we have found an optical shock which is the counterflow to the previously discovered ``northwest outflow'' from \lkha{}324SE.
\end{abstract}


\section{Introduction}

\begin{figure}[!htb]
\begin{center}
\plotone{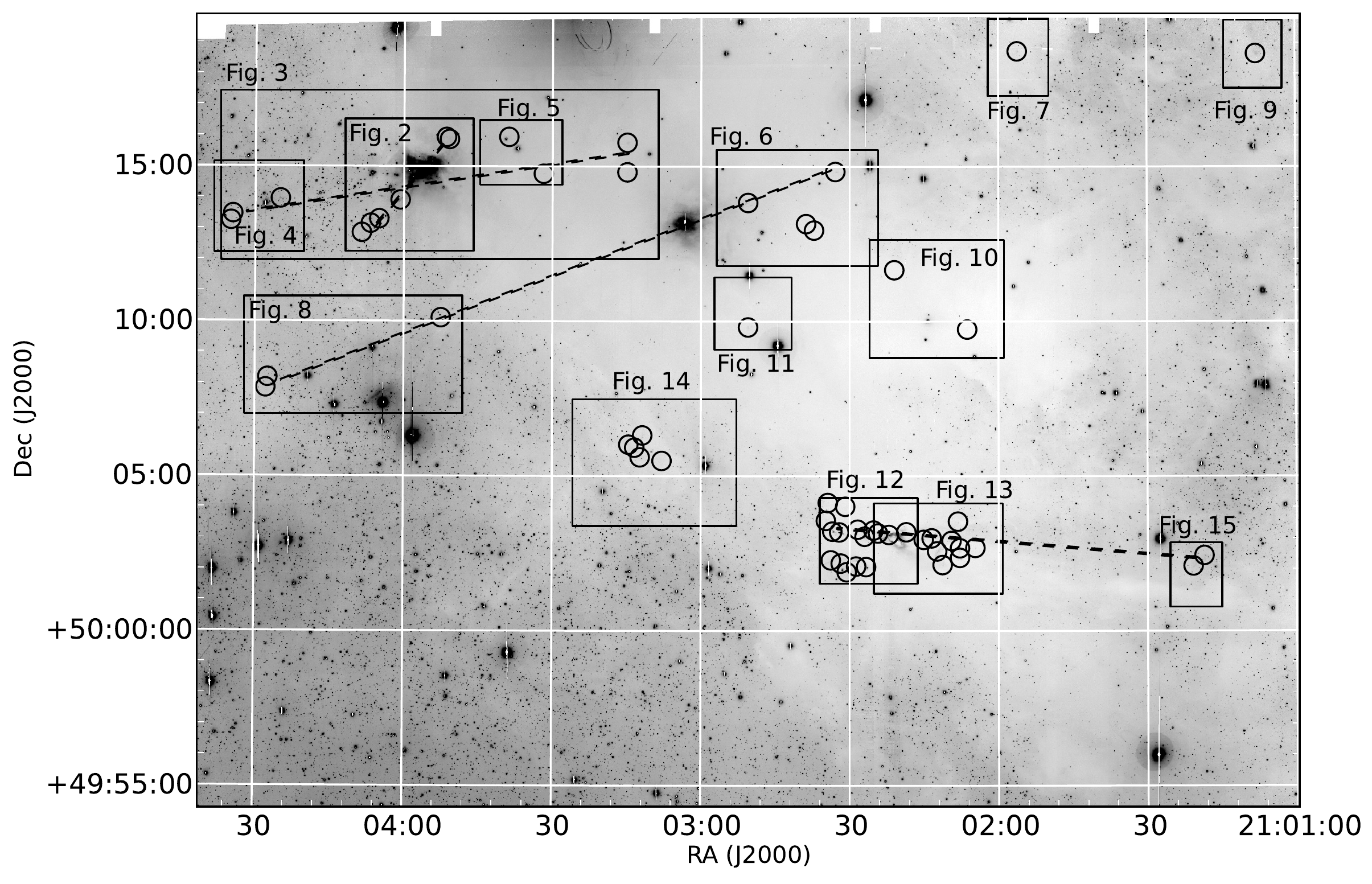}
\caption{\ha{} image of the L988 cloud HH objects are marked with circles and candidate giant flow axis are marked with dashed lines.  Boxes mark the areas shown in subsequent figures.}
\label{OverviewFig}
\end{center}
\end{figure}

Star formation is a dynamic process whereby newborn stars interact with their parent cloud.  High velocity outflows from accreting stars collide with parent molecular material generating shocks (known as Herbig-Haro objects when detected optically), opening cavities \citep{Quillen2005}, and driving turbulence \citep{MieschBally1994,Bally1999,ArceGoodman2002,Walawender2005}.  Shocks from outflows heat, dissociate, and ionize the gas.  They also inject kinetic energy and momentum into the cloud which may affect the rate of gravitational collapse of cores within these clouds \citep[e.g]{Leorat1990}.  Outflows may play a fundamental role in the evolution of star forming molecular clouds, turbulence generation, and cloud destruction.  In this paper, we have searched the Lynds~982, 984, and 988 dark cloud complex (L988 hereafter; see Fig.\ \ref{OverviewFig}) for shocks from protostellar outflows using tracers in both optical (\ha{} and \sii{}) and near-IR (\mh{}) tracers.  

L988 lies in Cygnus near the Cygnus OB7 association which is among the nearest of the Cygnus OB associations at roughly 740-800 pc.  L988 is part of a larger cloud complex known as Kh~141 \citep{Chavtasi1960}, or TGU~541 \citep{Dobashi2005}, and is sometimes called “The Northern Coalsack”.  The two regions of highest extinction within the Kh~141 complex are L988 and L1003 \citep[see][Fig.\ 28]{ReipurthSchneider2008}, both of which are active regions of star formation.

Distance estimates for the L988 complex range between 500 and 780~pc.  \cite{Chavarria81} estimated the distance at 700~pc based on photometry of several stars in the region.  Later, \cite{ChavarriadeLara81} estimated a distance of 780~pc.  \cite{Shevchenko91} found a distance of 550~pc based on extinction estimates.  \cite{Alves98} studied extinction toward the nearby L977 cloud and found a distance of 500~pc.  For calculations in this paper, we assume an intermediate distance of 600~pc.

The first outflow study in the region was a millimeter CO line survey by \cite{Clark86} who found four molecular outflows around IRAS sources which he designated $a$, $c$, $e$, and $f$.  Subsequently the flow surrounding IRAS~21007+4951 (\citeauthor{Clark86} source $a$) was imaged by \cite{Hodapp94} in the near-IR (K$'$) and by \cite{StaudeElsasser93} in r.  \cite{StaudeElsasser93} found four HH objects which were not assigned catalog numbers.  These HH objects correspond to our HH~1050 knots B, C, E, \& F (see \S\ref{Results}).

\cite{Felli92} searched for H$_2$O masers around young stars and found a maser associated with \citeauthor{Clark86} source $a$.  Their search did not detect masers coincident with \citeauthor{Clark86} sources $e$ or $f$.

\cite{HerbigDahm06} examined the \lkha{}324 region in L988 using broadband optical and near-IR imaging and optical spectroscopy and discovered a small cluster of YSOs surrounding \lkha{}324.  They found the age of the cluster surrounding \lkha{}324 to be 0.6-1.7~Myr depending upon the evolutionary model used in the analysis.  \cite{HerbigDahm06} also examined the \lkha{}324SE star (IRAS~21014+5001, \citeauthor{Clark86} source $c$) in detail using Keck HIRES spectroscopy.  They found features which they designate the ``northwest outflow'' (later designated HH~899) which is composed of several condensations prominent in \sii{} and \oii{} lines.  In a 7\arcsec{} long slit oriented at P.A.~=~129/309\arcdeg{}, they found three \sii{} condensations to the northwest at velocities of -160 to -185~\kms{} relative to the -18~\kms{} rest velocity of the star.  No red-shifted counterparts to the ``northwest outflow'' knots were detected.

\cite{Allen08} examined the cluster surrounding \citeauthor{Clark86} source $e$ with the Spitzer Space Telescope and cataloged young stars in the region.

\section{Observations}
Near infrared data for this project were obtained on the nights of 2006 July 11-13 on the United Kingdom Infrared Telescope (UKIRT) using the Wide Field InfraRed Camera (WFCAM, \citealt{Casali01}), which is comprised of four Rockwell Hawaii-II 2048$\times$2048 pixel arrays separated by 94\% of the size of an individual chip.  The instantaneous field of view is 0.21 square degrees, however to obtain a contiguous field of view, four pointings of the telescope must be used to fill in the space between arrays.  A four pointing ``tile'' covers approximately 0.8 square degrees.  In the J, H, and K filters, we obtained a total integration time of 6 minutes over the L988 tile.  In the \mh{} filter, we obtained 72 minutes of integration time.  WFCAM data were pipeline processed by the Cambridge Astronomical Survey Unit (CASU).  The 16 resulting image stacks were then mosaiced together using the Image Reduction and Analysis Facility\footnotemark (IRAF) to form the full field of view of the tile.

\footnotetext{IRAF is distributed by the National Optical Astronomy Observatories, which are operated by the Association of Universities for Research in Astronomy, Inc., under cooperative agreement with the National Science Foundation.}

Visible wavelength narrowband images were obtained on the night of 2006 May 28 on the Subaru Telescope using the Suprime-Cam instrument \citep{Miyazaki02}.  Suprime-Cam is a wide field prime focus camera comprised of ten 2048$\times$4096 pixel CCDs.  The instantaneous field of view is approximately 34\arcmin{}$\times$27\arcmin{}.  Images were taken in the \ha{} and \sii{} filters, each with a total exposure time of 50 minutes.

Subaru data were processed using IRAF's \verb'mscred' package.  Images were overscanned, trimmed, bias subtracted, and then flat fielded (using both dome and twilight flats) by the \verb'ccdproc' task.  Images were intensity matched using \verb'mscimatch' and stacked using \verb'mscstack' based on world coordinate system fits generated by \verb'msccmatch'.

Visible wavelength SDSS~i' images were obtained on the night of 2010 Aug 25 on the University of Hawaii 88 inch Telescope using the Wide Field Grism Spectrograph 2 (WFGS2; \citealt{Uehara04}) instrument in imaging mode.  WFGS2 uses the Tek2k CCD camera with 2048$\times$2048 pixels.  With the WFGS2 focal reducer, the field of view is approximately 11 arcminutes on a side.  SDSS~i' images were only obtained for the field centered on the IRAS~21007+4951 reflection nebula.  A total of 35 minutes of integration time was obtained.  Reductions, alignment, and stacking were performed using the \verb'ccdproc' package in IRAF.

\section{Results}\label{Results}
The L988 cloud has several active protostellar outflows which are visible in our narrowband images.  Some appear to be associated with cavities in the cloud which have been carved out by the action of the outflow.  Fig.\ \ref{OverviewFig} shows the region of the cloud covered by our \ha{} and \sii{} images.  Table \ref{TableShocks} contains a list of all Herbig-Haro objects (HH objects; \citealt{hhcat}) and molecular hydrogen shocks (MHO objects; \citealt{Davis2010}) in our field.  In the following paragraphs, we discuss individual shocks, organized into proposed outflow groups

\begin{figure}[!htb]
\includegraphics[width=3in]{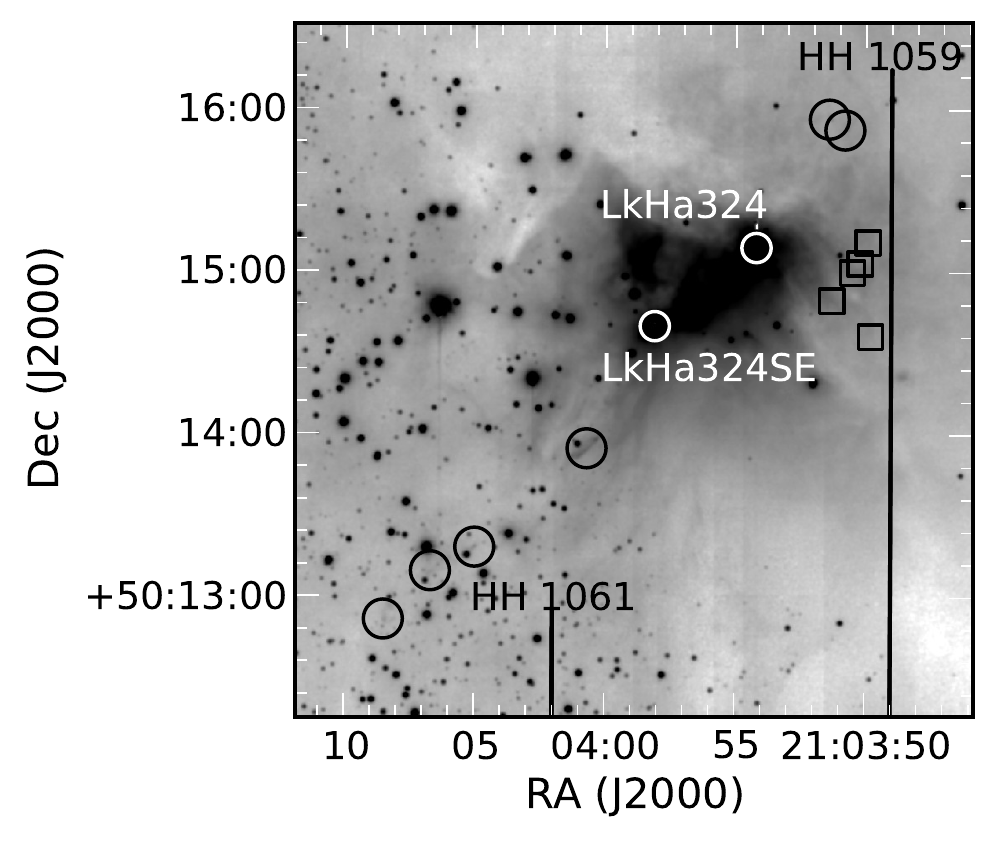}
\caption{An \ha{} image of the \lkha{}324 region (IRAS~21023+5002; \citeauthor{Clark86} source $e$) region which includes the HH~1061 and 1059 shocks.  \lkha{}324 and \lkha{}324SE are embedded in the reflection nebula and their positions are marked with white circles.  The IRAS source is coincident with \lkha{}324SE.  Squares mark the positions of knots in the MHO~954 shock complex.}
\label{FigHH_1059_1061}
\end{figure}

\noindent {\bf HH~1061 \& 1059:}  
HH~1061 (Fig.\ \ref{FigHH_1059_1061}) lies 0.4\arcmin{} south-southeast of \lkha{}324SE (\citeauthor{Clark86} source $e$, IRAS~21023+5002; see Table \ref{TableSources}) and its associated reflection nebula and cluster of sources.  The primary component is a thin \ha{} bright filament, oriented northwest-southeast and roughly 10\arcsec{} long.  There are additional faint filaments of emission (knots B, C, \& D) stretching about 1.3\arcmin{} further to the southeast.  The shock lies in a low extinction region which may be a cavity outflows have blown out of the cloud.

There are two compact knots (HH~1059A \& B) in both the \ha{} and \sii{} images about 2\arcmin{} to the northwest of HH~1061 (Fig.\ \ref{FigHH_1059_1061}).  They lie along the line defined by the HH~1061 knots and lie across the cluster from HH~1061, thus they are likely counterflow components to HH~1061.

There are several young stars catalogued by \cite{Allen08} in the \lkha{}324 cluster, therefore it is not possible to determine with confidence which of the young stars may drive this outflow.  However, using spectroscopy \cite{HerbigDahm06} detected a high velocity ``northwest outflow''  (HH~899) within 3\arcsec{} of the \lkha{}324SE star.  That source lies close to the line defined by HH~1061 knots and HH~1059.  The high velocity knots of \cite{HerbigDahm06} were detected in a spectrograph slit oriented at P.A.$\sim$129/309\arcdeg{}, the HH~1061/1059 flow lies at a position angle of $\sim$151/331\arcdeg{}.  Given that the slit may not have been placed precisely along the outflow axis (which was unknown at the time of the observation), these values seem to be in reasonable agreement and it is likely that \lkha{}324SE is the driving source and that HH~899 is part of a flow with HH~1061 \& HH~1059.

The \lkha{}324SE source has been detected in several infrared surveys (see Table \ref{TableSources}).  Those measured fluxes were used to fit the spectral energy distribution (SED) using the models of \cite{Robitalle07} (see Fig.\ \ref{RobitalleModels}a).  The SED shows evidence of strong disk and envelope components.

\cite{Clark86} found a bipolar outflow surrounding \lkha{}324SE.  The outflow is oriented roughly North-South with the blue-shifted lobe to the North, moving into the higher extinction region.  It is unclear whether the CO outflow of \cite{Clark86} and HH~1061 \& 1059 trace the same flow.  While the \citeauthor{Clark86} outflow appears to be oriented North-South (P.A.~=~0/180\arcdeg{}), the low resolution in \citeauthor{Clark86} Figure 1 makes the association of the two unclear.

\noindent {\bf MHO~954:}
A bow shaped \mh{} knot (MHO~954A) lies $\sim$1\arcmin{} due west of \lkha{}324SE (see Fig.\ \ref{FigHH_1059_1061}) along with a line of fainter knots (MHO~954B-E).  Based on their position relative to the core of the cluster, it seems unlikely that these shocks are from the same outflow as HH~1061 and 1059.

\begin{figure}[!htb]
\includegraphics[width=6.5in]{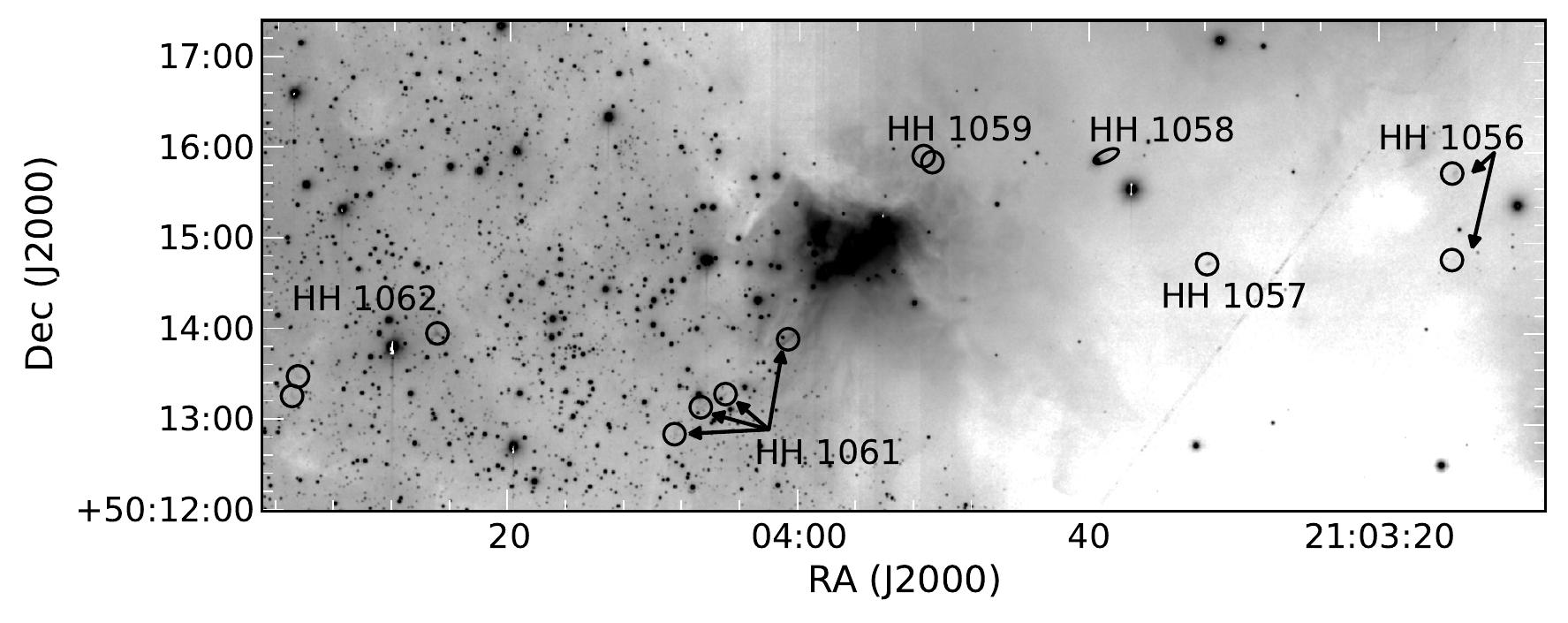}
\caption{An \ha{} image of the \lkha{}324 region (IRAS~21023+5002; \citeauthor{Clark86} source $e$) region which includes the shock HH~1056-1059 \& 1061-1062 shocks.}
\label{FigHH_1056_to_1062}
\end{figure}

\noindent {\bf HH~1057 \& 1056:}
These are compact, \ha{} bright shocks (Fig.\ \ref{FigHH_1056_to_1062}) which lie in a dark portion of the cloud.  There is no clear association with a known flow or source, however we note that the MHO~954 shock, HH~1057, and 1056~B lie along an axis which passes through the cluster of sources near \lkha{}324SE and thus may be a single outflow emerging from one of the sources embedded in the reflection nebula.

\begin{figure}[!htb]
\includegraphics[width=3in]{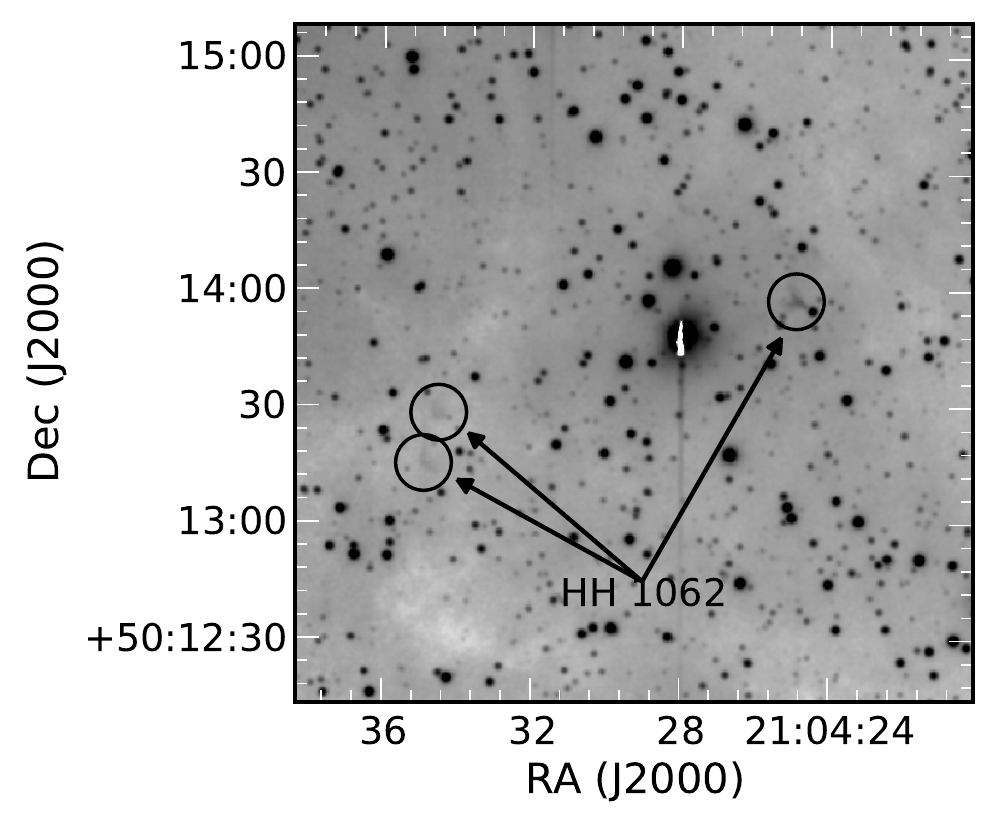}
\caption{An \ha{} image of the HH~1062 shock system.}
\label{FigHH_1062}
\end{figure}

\noindent {\bf HH~1062:}
These three knots lie in a low extinction region East of the cloud (Figs.\ \ref{FigHH_1056_to_1062} \& \ref{FigHH_1062}).  They surround a bright star visible in our narrowband \ha{} and \sii{} images which corresponds to an IRAS source (IRAS~21028+5001).  This star was also detected by WISE (WISE~J210428.01+501348.5) and it lies roughly on a line drawn between the shocks, but it is not clear if they are driven by it.  Fluxes for the star are listed in Table \ref{TableSources} and the model SED is shown in Fig.\ \ref{RobitalleModels}b.  

HH~1062~A lies 26\arcsec{} to the West of the source along PA$\sim$288\arcdeg{} while knots B and C lie 52\arcsec{} and 59\arcsec{} to the East along PA$\sim$107\arcdeg{} and 116\arcdeg{} respectively.  We also note that these shocks lie roughly along a line defined by the HH~1057 and 1056 shocks that passes through the reflection nebula.  HH~1062 may be a shock in a larger flow and is perhaps not driven by the nearby IRAS~21028+5001.

\begin{figure}[!htb]
\includegraphics[width=3in]{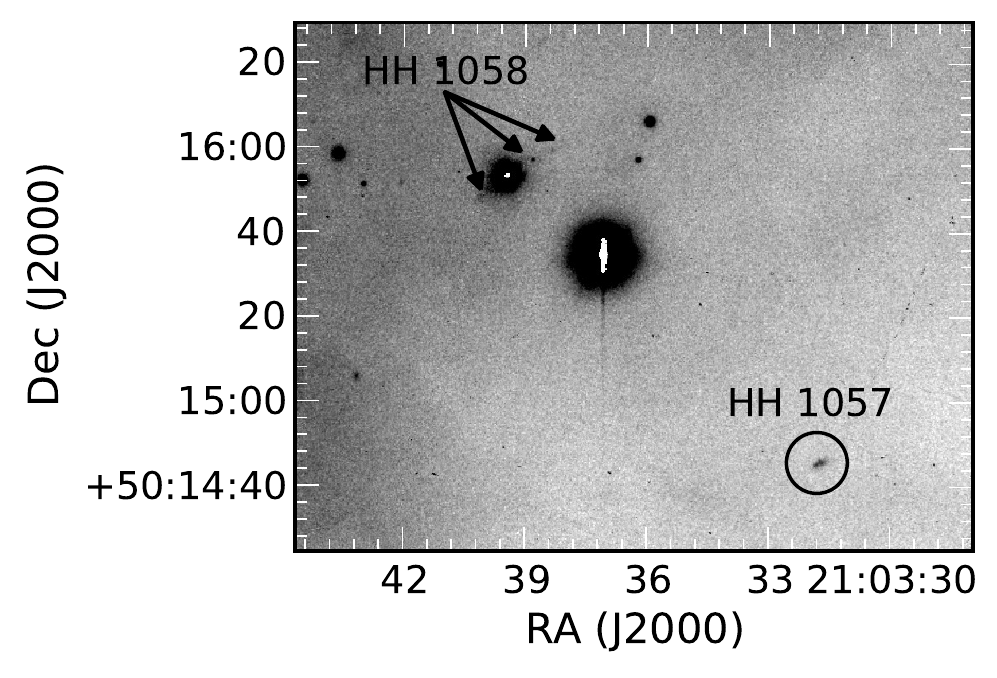}
\caption{An \ha{} image of the HH~1058 and 1057 shocks.}
\label{FigHH_1057_1058}
\end{figure}

\noindent {\bf HH~1058:}
This is a faint, \ha{} only shock (Figs.\ \ref{FigHH_1056_to_1062} \& \ref{FigHH_1057_1058}) which appears to be a jet emerging from an optically visible star along PA$\sim$125/305\arcdeg{}.  The star was detected by both 2MASS and WISE (WISE~J210339.46+501552.9) and those magnitudes are listed in Table \ref{TableSources}.  Using those fluxes, we fit the SED using the models of \cite{Robitalle07} (see Fig.\ \ref{RobitalleModels}c).  The best fit models show that this is a low mass young star ($\sim$0.25~\Msun{}) with a relatively low disk accretion rate ($\sim 10^{-11}$~\Msun{}/yr).

\begin{figure}[!htb]
\includegraphics[width=6.5in]{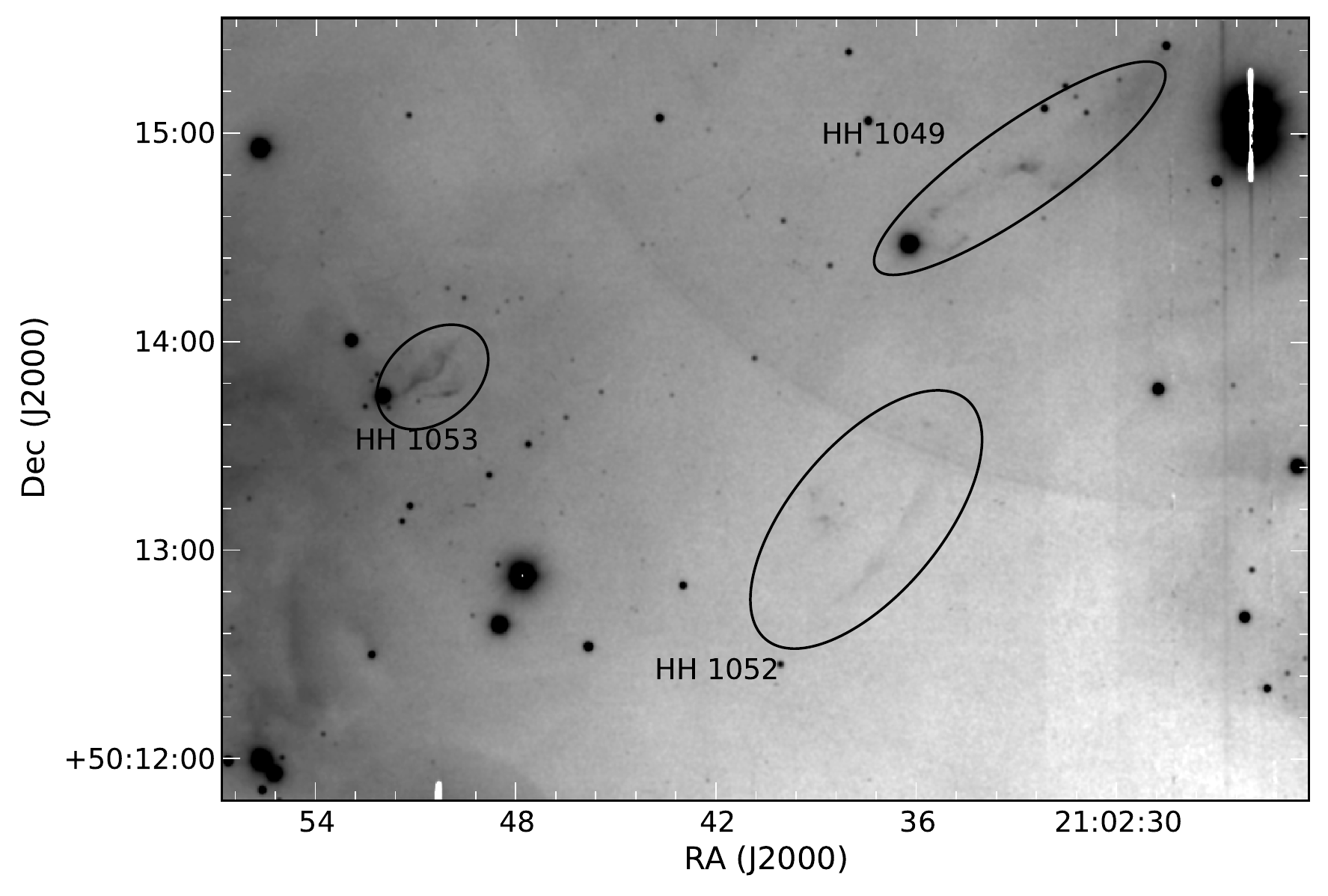}
\caption{An \ha{} image of the HH~1053, 1049, \& 1052 region.  The large reflection nebula on the left side of the image surrounds IRAS~21014+5001 (\citeauthor{Clark86} source $c$; see Fig.\ \ref{OverviewFig} for the relationship).}
\label{FigHH_1049_1052_1053}
\end{figure}

\begin{figure}[!htb]
\epsscale{0.7}
\includegraphics[width=3in]{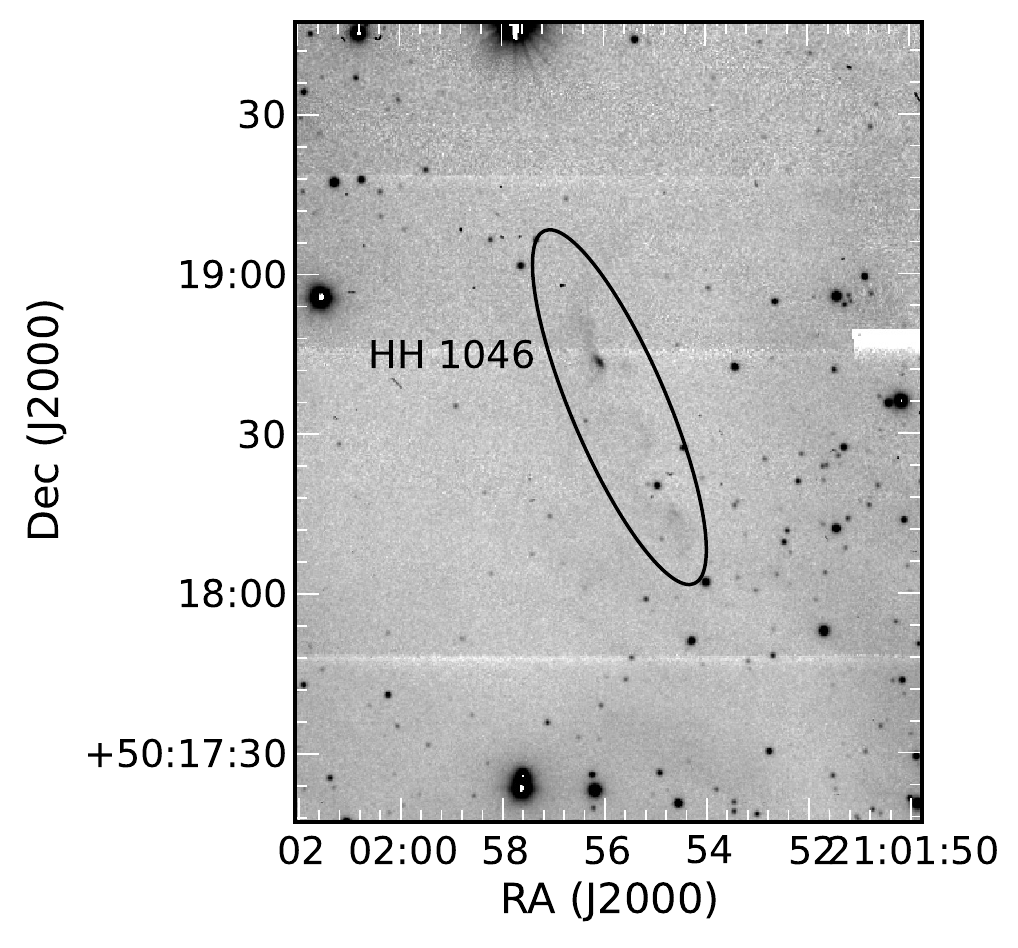}
\caption{A \sii{} image of HH~1046 which lies at the north edge of our Subaru field.}
\label{FigHH_1046}
\end{figure}

\noindent {\bf HH~1053, 1049, 1046, 1060, \& 1063:}  
The HH~1053, 1049 (Fig.\ \ref{FigHH_1049_1052_1053}), \& 1046 shocks (Fig.\ \ref{FigHH_1046}) appear to all be the western components of a single flow powered by IRAS~21014+5001 (aka \citeauthor{Clark86} source $c$), it is possible that HH~1049 or 1046 are from another source, but the alignment of these three shocks with IRAS~21014+5001 appears convincing.  

\begin{figure}[!htb]
\epsscale{1.0}
\includegraphics[width=6.5in]{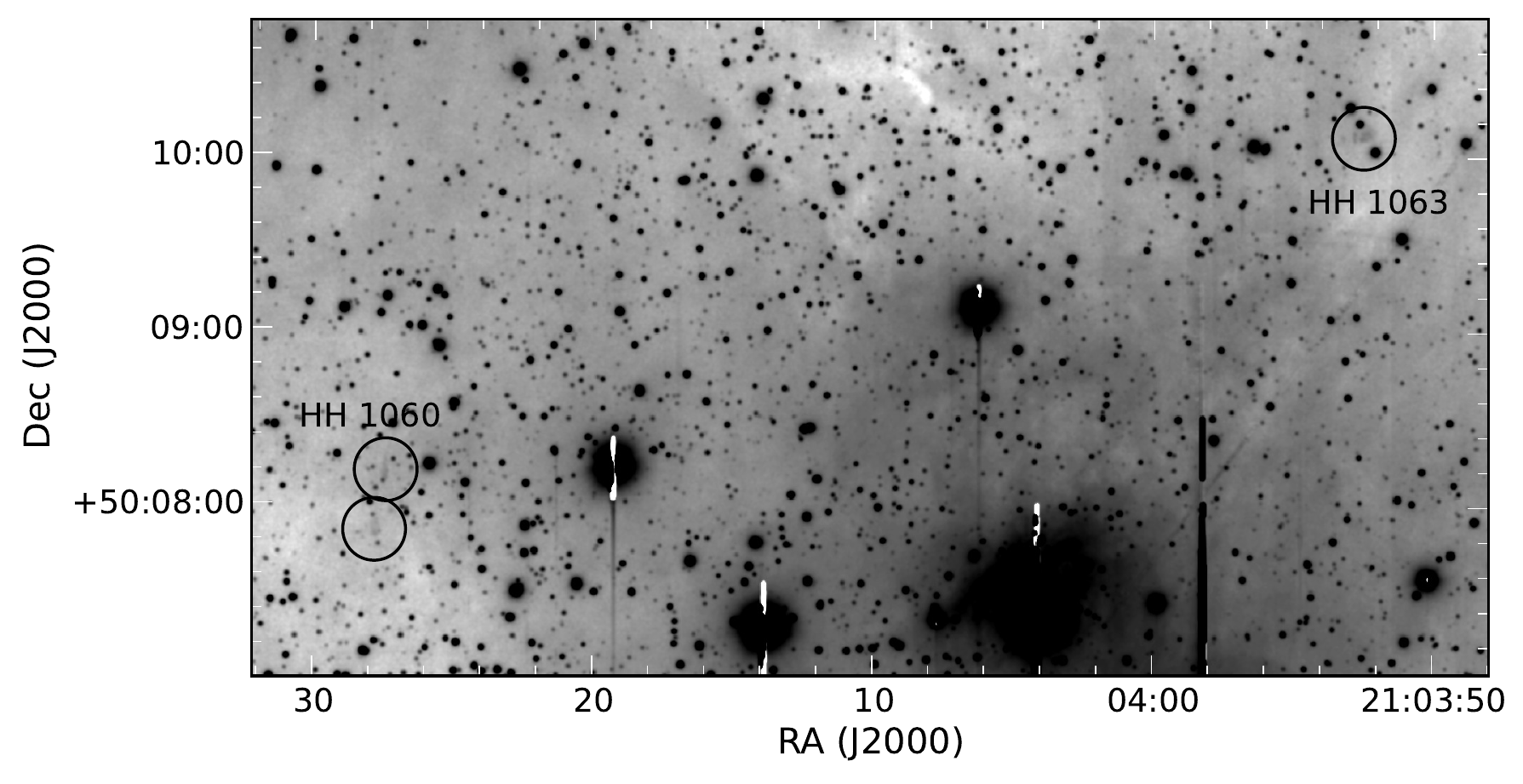}
\caption{An \ha{} image of the HH~1060 \& 1063 region.}
\label{FigHH_1060_1063}
\end{figure}

The corresponding counterflow to the southeast emerges into a low extinction region of the cloud and is composed of the HH~1060 and 1063 shocks (Fig.\ \ref{FigHH_1060_1063}).  If this is all one flow which emerges from that source, then the length of the flow is 26.5\arcmin{} which corresponds to a length of 4.6~pc at an assumed distance of 600~pc.

Clark designated IRAS~21014+5001 as a molecular outflow source based upon finding a patch of blue-shifted gas West of the source (see \citealt{Clark86} Fig.\ 1).  This patch of blue-shifted CO does not correspond to our HH~1053, 1049, 1046 outflow.  The blue-shifted CO is centered South of the axis defined by our HH objects.  Though the northernmost contour in \citet[Fig.\ 1]{Clark86} comes close to our flow axis near HH~1049, the bulk of the blue-shifted CO is closer to our HH~1052 (see below).  The IRAS~21014+5001 source was detected by WISE, the fluxes are listed in Table \ref{TableSources} and the model SED is shown in Fig.\ \ref{RobitalleModels}d.

\noindent {\bf MHO~955:}
There is a star visible in the J, H, \& K images (and faintly in \ha{} and \sii{}) 1.2\arcmin{} east-southeast of IRAS~21014+5001 near the axis defined by the HH~1053, 1049, 1046, 1060, 1063 outflow which is coincident with a 0.2\arcmin{} long \mh{} filament (MHO~955) pointing northeast from the star, at first glance it appears that it is emanating from the star.  That star was determined to be a Class~I protostar by \cite{Allen08}, however it lies along the axis of the HH~1053, 1049, 1046 flow from IRAS~21014+5001 and so may alternatively be a shock in the embedded counterflow which happens to be coincident with a star along our line of sight.

\begin{figure}[!htb]
\epsscale{0.7}
\includegraphics[width=3in]{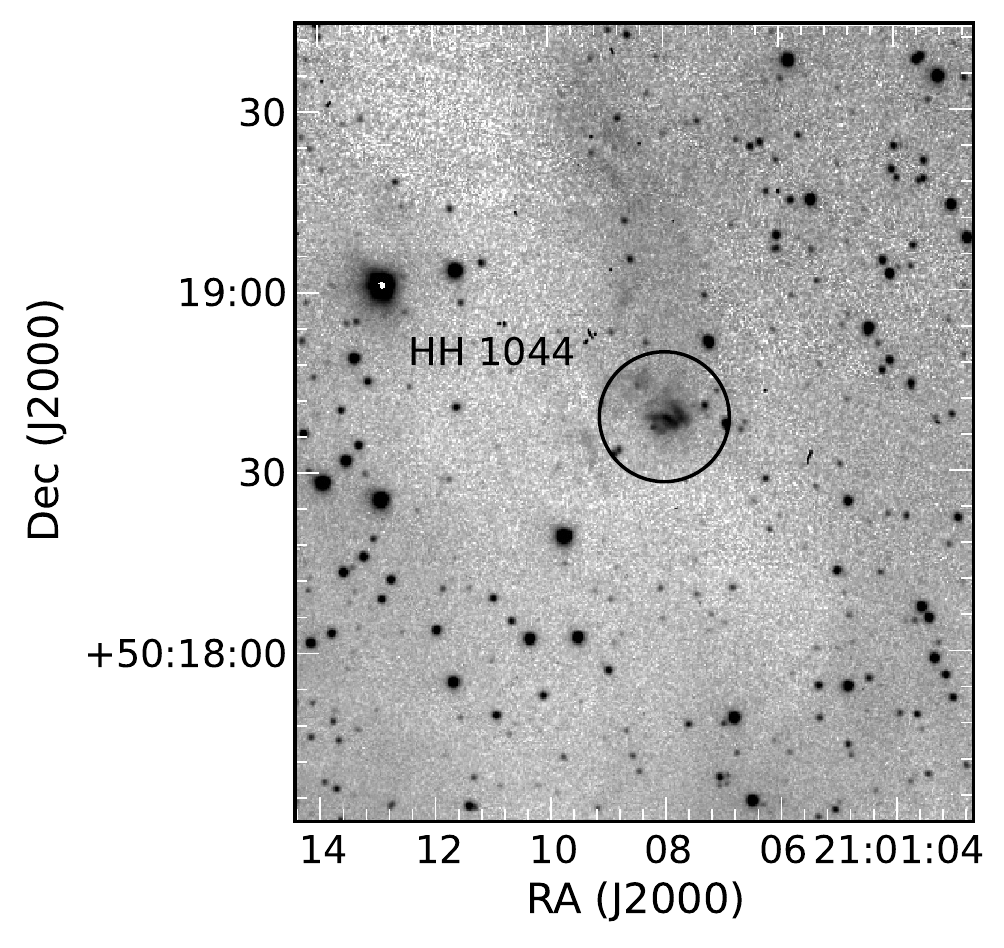}
\caption{A \sii{} image of the HH~1044 shock which lies near the extreme northwest edge of our Subaru field.}
\label{FigHH_1044}
\end{figure}

\noindent {\bf HH~1044:}  
This is a \sii{} bright knot (Fig.\ \ref{FigHH_1044}) which lies at the northwest corner of the Subaru image.  There appears to be a faint \ha{} filament extending due North from it, but this may also be an illuminated cloud edge.  IRAS~20595+5009 lies 2.5\arcmin{} north of the shock, outside of the field of view of our \ha{} and \sii{} images.

\noindent {\bf HH~1052:}
The A \& B knots of this object (Figs.\ \ref{FigHH_1049_1052_1053}) are a pair of faint \ha{} and \sii{} filaments.  The B component has an \mh{} counterpart (MHO~956).  There is a faint V-shaped reflection nebula visible in the H and K images 0.3\arcmin{} southeast of the knots.  The reflection nebula opens toward both of the knots.  At the apex of the reflection nebula is a candidate for the source star.  While it is not detected in our J, H, or \Ks{} images, it is detected by WISE (WISE~J210240.13+501236.5; Table \ref{TableSources}).  In addition, there is another candidate source star visible in our near-IR images which lies 19\arcsec{} to the southeast.  It is coincident with IRAS~21010+5000 and is detected in the WISE catalog (WISE~J210242.41+501227.8; Table \ref{TableSources}; Fig.\ \ref{RobitalleModels}e).  

The blue-shifted CO which was discovered by \cite{Clark86} is coincident with the HH~1052 shock system.  \citeauthor{Clark86} associated this with IRAS~21014+5001, however we associate that source with the HH~1053, 1049, 1046 flow.

\noindent {\bf HH~1047:}  
This shock system appears to be a curved, C-shaped outflow.  The C-shaped curve suggests that the source is moving to the southeast \citep{BallyReipurth2001}.

In our near-IR images, several stars (many with corresponding WISE detections) lie on or near the arc of the flow and would be source candidates.  Of these, three stand out as being directly along the arc of the flow (WISE~J210214.15+501013.0, WISE~J210218.93+501102.3, \& WISE~J210217.70+501046.5).  The first of these is optically visible while the other two are visible only in our near-IR images.  We examined the models of \cite{Robitalle07} built based upon the WISE and 2MASS magnitudes of each of these and have selected the optically visible star (WISE~J210214.15+501013.0) as the best candidate as it is the only one of the three with a significant envelope flux in the fitted SED.  The fluxes for this star are listed in Table \ref{TableSources} and the result of the \cite{Robitalle07} model fit can be seen in Fig.\ \ref{RobitalleModels}f.

\begin{figure}[!htb]
\includegraphics[width=3in]{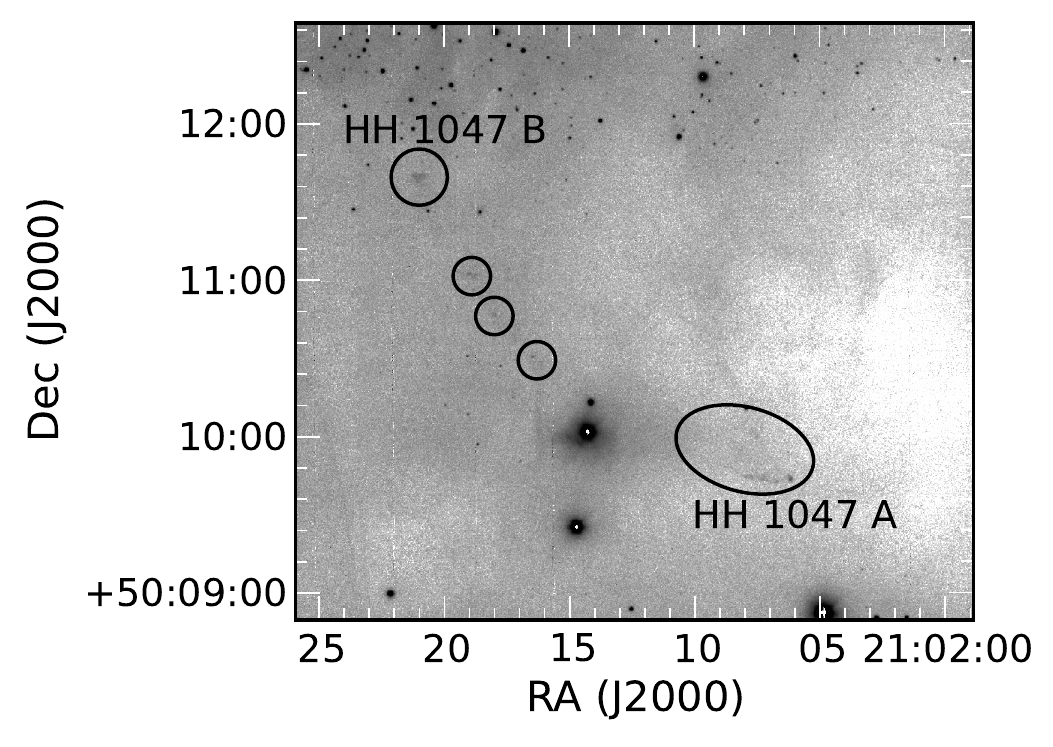}
\caption{A \sii{} image of the HH~1047 outflow.  The WISE source discussed in the text is labeled.}
\label{FigHH_1047}
\end{figure}

It is also possible that this is merely a bow shock driven by a more distant source, however, the morphology is suggestive of a jet and even if this were a bow shock, the arc of the curve would indicate a forward facing bow shock coming from the northwest and there are no strong flows visible in our images in that direction.  The star V1331 Cyg and its associated outflow (HH~389) lie along that axis a few arcminutes outside of the field of view of our SuPrimeCam images.  The \sii{} images of \cite{MundtEisloffel1998} show shocks to the north and south of V1331 Cyg.  Their orientations relative to the source are inconsistent with HH~1047 being a component in that flow.

\begin{figure}[!htb]
\includegraphics[width=3in]{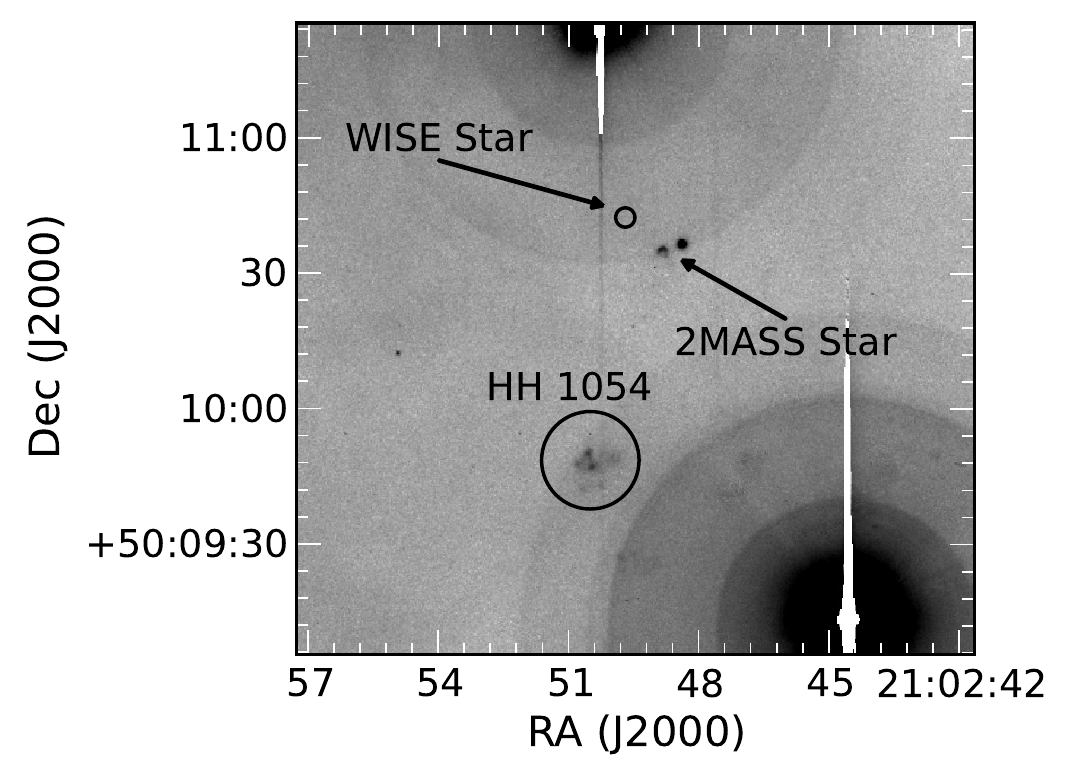}
\caption{A \sii{} image of the HH~1054 shock.  The locations of 2MASS 21024889+5010351 (visible in the \sii{} image) and WISE J210249.68+501041.9 (only visible in the near-IR) are marked.}
\label{FigHH1054}
\end{figure}

\noindent {\bf HH~1054:}  
This is a compact cluster of \sii{} dominant knots (Fig.\ \ref{FigHH1054}) coincident with a short 7\arcsec{}  V-shaped \mh{} shock (MHO~957) which opens to the North.  In that direction, there are two optical and near-IR stars about 1 arcminute away, both with reflection nebulosity.  The first source (2MASS 21024889+5010351; Table \ref{TableSources}) is seen in our visible light narrowband images as well and has a small reflection nebula surrounding it which (in our optical images) appears to open southward.  This source has a brighter companion star 3.5\arcsec{} to the West (WISE J210248.45+501036.1).  This star is a likely candidate for the outflow source as the orientation of the reflection nebula is aligned with the presumed flow direction.

Another candidate source star with a reflection nebula (WISE J210249.68+501041.9; Table \ref{TableSources}) is visible in our near-IR images and lies about 1 arcminute North of the \mh{} shock.  Its reflection nebula, however, appears to open to the East.  The source star does not appear in the 2MASS catalog.  We were unable to determine reliable J, H, and K magnitudes for this star due to the presence of the reflection nebula, but it is clearly very red (invisible in J, faint in H, and bright in K).

\begin{figure}[!htb]
\begin{center}
\includegraphics[width=6.5in]{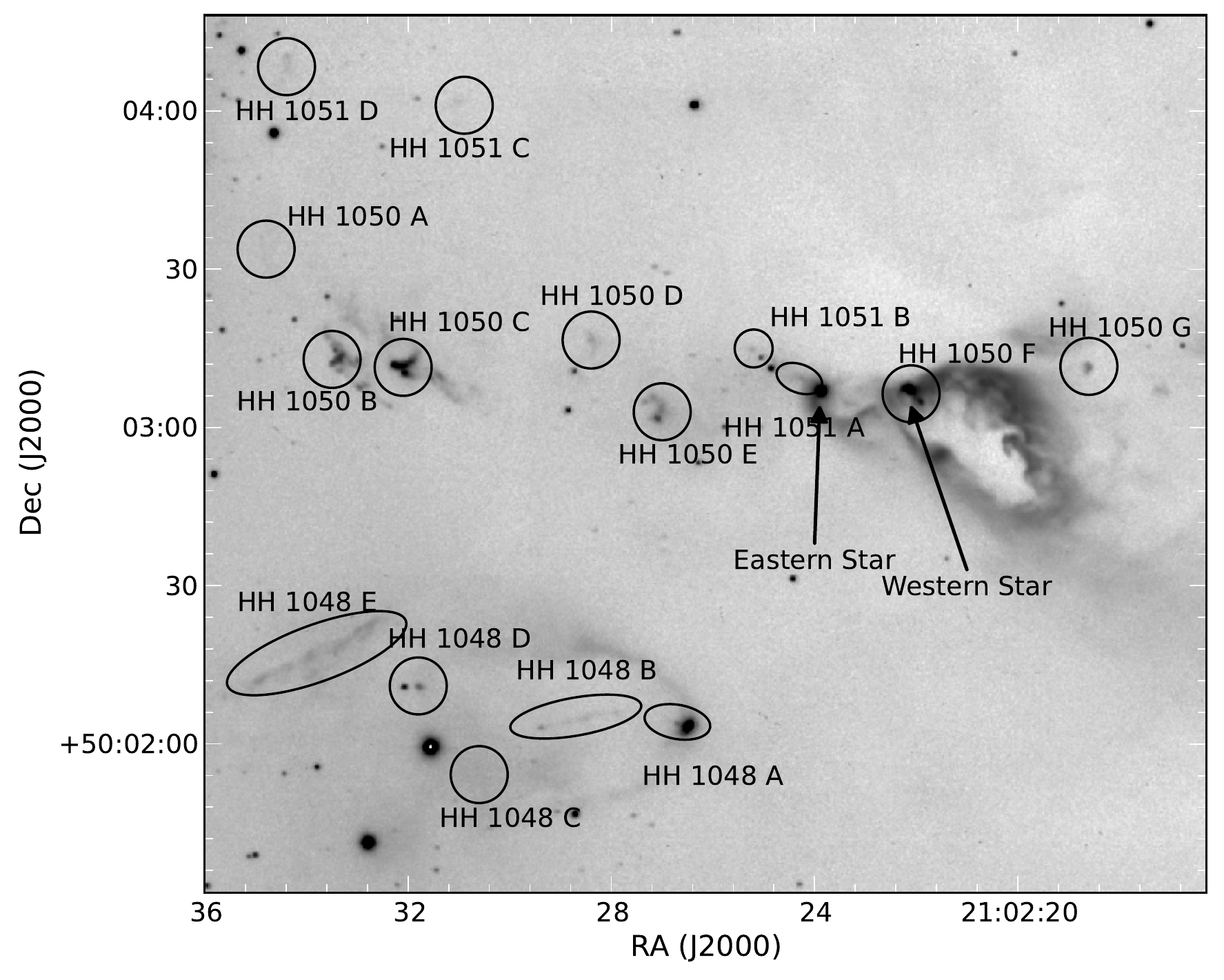}
\caption{A \sii{} image of the area around the HH~1048, 1050, and 1051 shock systems.  The positions of the Eastern and Western stars described in the text are indicated with arrows and the circles mark the positions of HH objects.}
\label{FigHH_1048_1050E_1051}
\end{center}
\end{figure}

\noindent {\bf HH~1051:}  
The HH~1050 and 1051 outflow system is a complex arrangement of shocks and reflection nebulosity (Fig.\ \ref{FigHH_1048_1050E_1051}).  The reflection nebula appears to outline a two-lobed cavity surrounding \citeauthor{Clark86} source $a$.

There are two stars embedded in the reflection nebula.  The western star (WISE J210222.70+500308.3; Table \ref{TableSources}; Fig.\ \ref{RobitalleModels}g) is not readily seen in the visible light images, but is detectable in the J images and is increasingly bright at the longer (H and K) wavelengths.  The eastern star (WISE J210223.85+500306.8; Table \ref{TableSources}; Fig.\ \ref{RobitalleModels}h) lies 15\arcsec{} East of the western star, is visible at all wavelengths, and is brighter than the western star at i', J, H, and K.  

\begin{figure}[!htb]
\begin{center}
\includegraphics[width=6.5in]{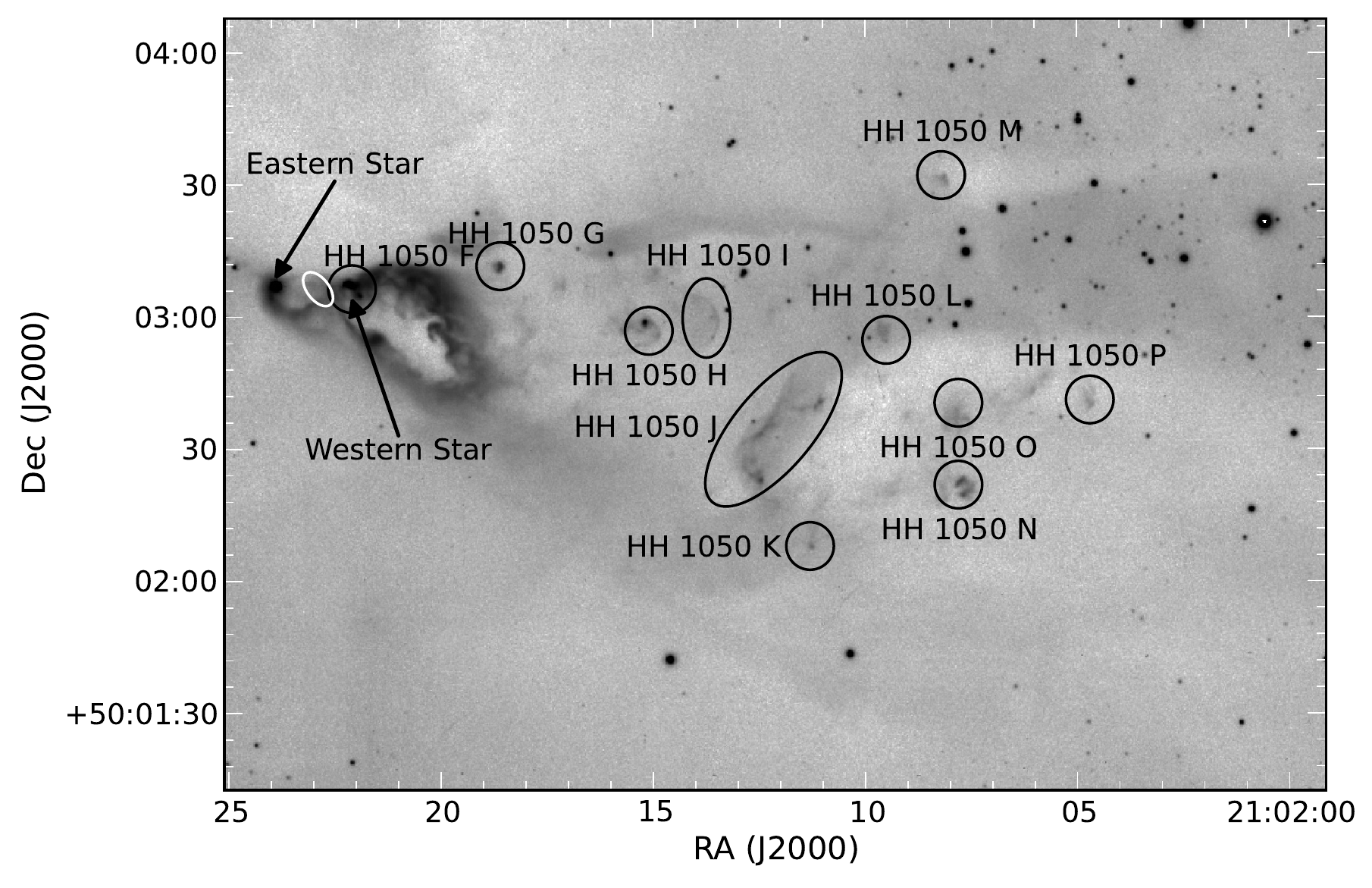}
\caption{An \ha{} image of the western half of the HH~1050 shock system.  The positions of the Eastern and Western stars described in the text are indicated with arrows and the circles mark the positions of HH objects.  The error ellipse for the position of IRAS~21007+4951 is indicated in white.}
\label{FigHH_1050W}
\end{center}
\end{figure}

IRAS~21007+4951 lies in between the Eastern and Western stars, slightly closer to the Western star.  Neither star lies within the error ellipse of the IRAS source (Fig.\ \ref{FigHH_1050W}).  It is possible that the IRAS source is a blend of the two WISE sources.

The eastern star appears to drive a faint, highly collimated \sii{} jet (HH~1051~A \& B; see Fig.\ \ref{FigHH_1048_1050E_1051}) along position angle 60\arcdeg{}.  To the northeast along the axis defined by the jet, there are two faint shocks (HH~1051~C \& D) which lie 88\arcsec{} and 119\arcsec{} from the source respectively.  These faint shocks are just visible in the \sii{} images and have no corresponding signal in the \ha{} or i$'$ images, nor are there faint stars at those locations in any of the near-IR images.  The southwestern portion of the jet blends with the southern edge of the reflection nebula.

Further to the southwest are faint shocks (see Fig.\ \ref{FigHH_1050W}) which may be part of this flow, however they overlap with the western lobe of the HH~1050 flow described below.

\noindent {\bf HH~1050 \& 1045:}  
The western of the two stars in the region appears to drive a long outflow (HH~1050; see Fig.\ \ref{FigHH_1050W}) in the East-West direction.  This outflow is much more extended than the HH~1051 outflow and lies along a nearly East-West line (PA$\sim$90\arcdeg{}/270\arcdeg{}).  There is also a bright, compact \mh{}, \ha{}, and \sii{} knot (HH~1050~F / MHO~958) 4\arcsec{} west of the western star.  

Knots A-E of this shock system (Fig.\ \ref{FigHH_1048_1050E_1051}) are the eastern component of the flow.  The furthest is knot A which lies 1.6\arcmin{} east of the source.  The B \& C knots are bright in both \ha{} and \sii{} and both have \mh{} counterparts (MHO~958).  Knots D \& E are fainter \ha{} and \sii{} knots with an \mh{} knot (MHO~958) a few arcseconds southwest of E.

HH~1050 knots F-P appear to comprise the western half of the outflow.  The innermost shocks (knots F and G) appear to have the morphology of compact bow shocks emanating from the area at the center of the reflection nebula.

Knot J appears to be a shock outlining the edge of a dense cloud edge which is visible in the \ha{} and \mh{} images.  The southern knots (J, K, N, O, and P) could be distant components of HH~1051, or the southern half of a less collimated component of HH~1050.  Because there is faint, filamentary emission which appears to connect many of the shocks in this region, we favor the explanation that all of the knots (F-P) are part of the HH~1050 outflow, but the possibility that at least some of these (especially the most southern knots J, K, and N) are associated with the HH~1051 flow cannot be excluded.

\begin{figure}[!htb]
\includegraphics[width=3in]{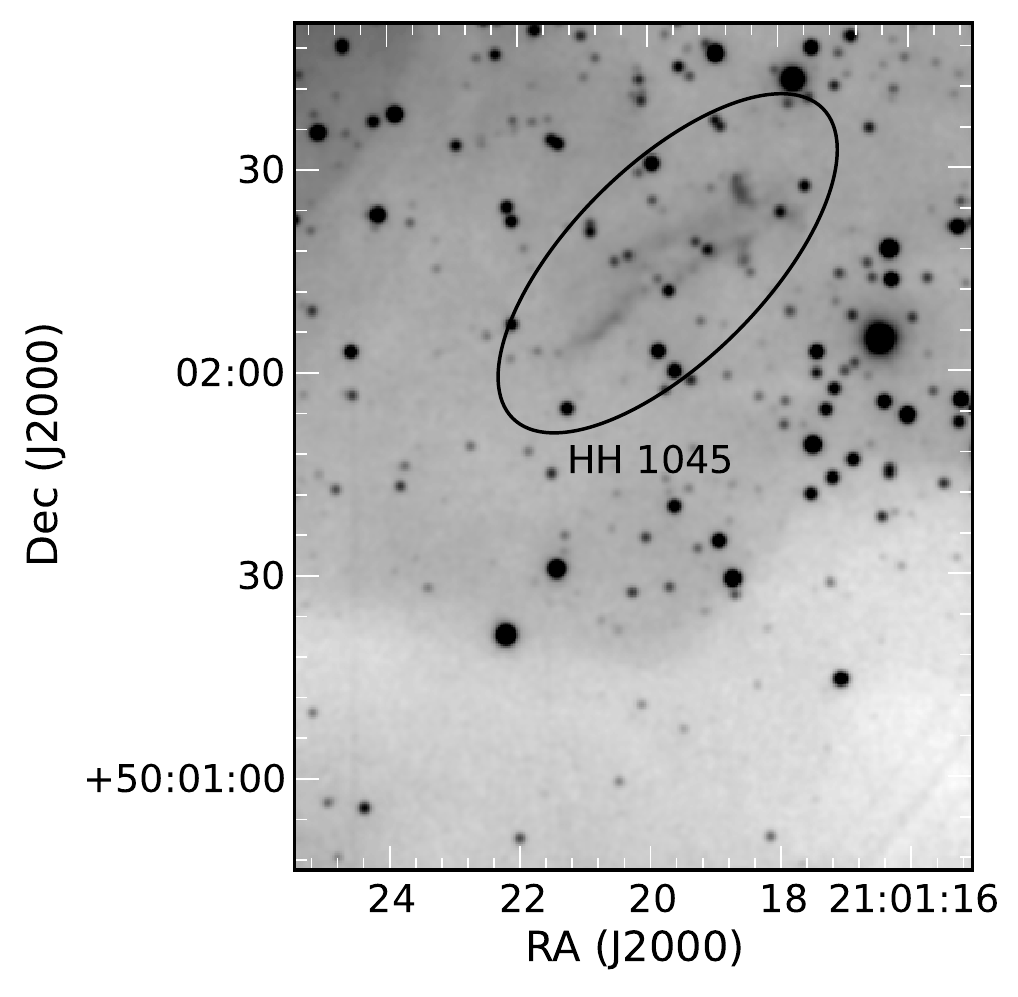}
\caption{A \sii{} image of the HH~1045 shock, the western end of the giant IRAS~21007+4951 outflow.}
\label{FigHH_1045}
\end{figure}

The pair of knots which make up HH~1045 comprise a $\sim$30\arcsec{} long filament of \ha{} \& \sii{} emission which lies 8.1\arcmin{} West of the HH~1050 outflow source along the axis of the flow (see Fig.\ \ref{FigHH_1045}).  This system appears to be a distant bow shock in that outflow.  At an assumed distance of 600~pc, that makes the total length of the HH~1050 outflow 1.7~pc.

The reflection nebula surrounding the HH~1050 outflow source outlines a V-shaped cavity about 2\arcmin{} long with a wide $\sim$45\arcdeg{} opening angle (Fig.\ \ref{FigHH_1050W}).  The northern half of this cavity, past knots L, \& K, emerges into a low extinction region in which background stars can be seen.  The southern half of the cavity is a higher extinction region.

The \cite{Clark86} outflow map shows red- and blue-shifted lobes which correspond to our HH~1050 outflow.  The eastern half of the flow (knots A-E) correspond to red-shifted CO and the western half (knots F-P) correspond to blue-shifted CO.

\noindent {\bf HH~1048:}  
Knot A of HH~1048 is a small, compact blob of emission just emerging from a star which lies at the apex of a C-shaped reflection nebula (Fig.\ \ref{FigHH_1048_1050E_1051}) and which is coincident with WISE J210226.45+500203.4 (Table \ref{TableSources}).  Knot B of HH~1048 is an \ha{} and \sii{} filament which lies in the middle of that C-shaped reflection nebula. 

HH~1048 knot C is a compact, \ha{} only bow which lies near the eastern end of the reflection nebula, about 0.5\arcmin{} from the star.  HH~1048~D is a compact, nearly starlike knot of \ha{} and \sii{} emission.  HH~1048 knot E is another filament which lies roughly 1\arcmin{} east of the star and slightly north of the line defined by the knot B filament, it may trace the northern edge of the outflow cavity.  The J, H, \& K images also show one arc of the reflection nebula extending roughly 15\arcsec{} northeast of the source star.

The \cite{Clark86} outflow map shows a blob of blue-shifted CO which is connected to the blue-shifted lobe of his source $a$.  While \citeauthor{Clark86} did not discuss this as a separate outflow, it is perfectly coincident in position with our HH~1048 knots A \& B and appears to match the outline of the associated reflection nebula in size.

\begin{figure}[!htb]
\includegraphics[width=3in]{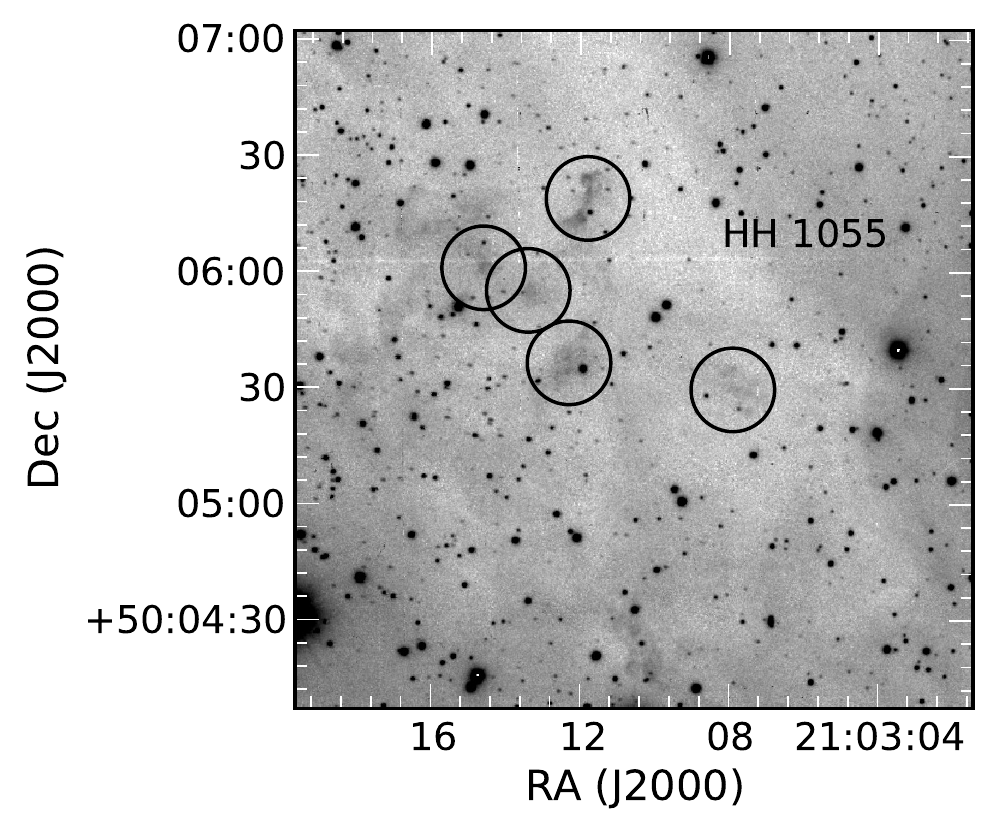}
\caption{A \sii{} image of the HH~1055 shock, a possible eastern component of the giant IRAS~21007+4951 outflow.}
\label{FigHH_1055}
\end{figure}

\begin{figure}[!htb]
\plotone{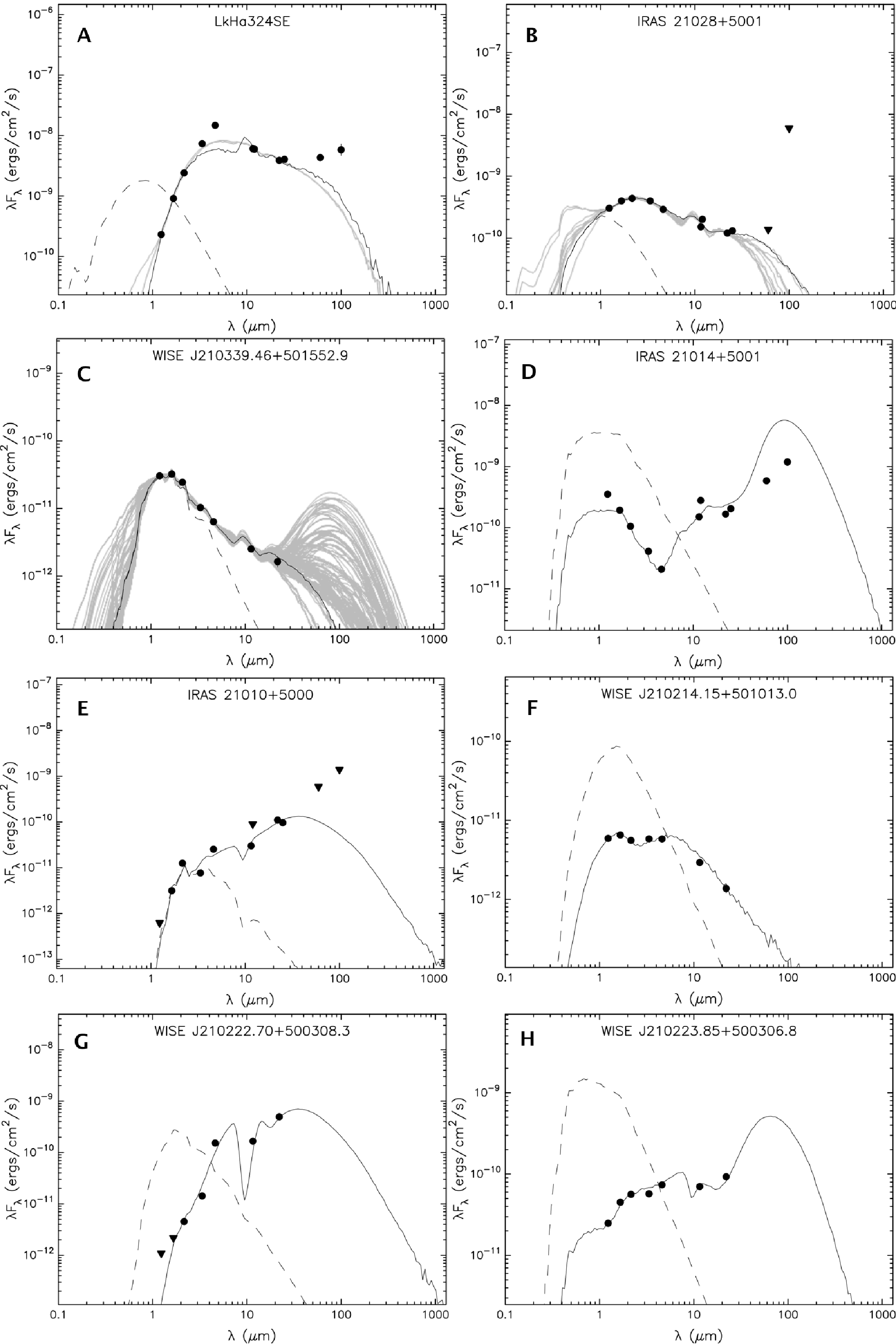}
\caption{Spectral energy distributions based on 2MASS, WISE, and IRAS catalog data fit with models by \cite{Robitalle07}.}
\label{RobitalleModels}
\end{figure}

\noindent {\bf HH~1055:}
This shock system is a cluster of faint, \sii{} bright, low surface brightness shock filaments (Fig.\ \ref{FigHH_1055}) embedded in some patchy dust filaments along the eastern edge of the cloud.  They lie near to the axis defined by the HH~1051 and 1050 outflows from the IRAS~21007+4951 source, however a positive association with those outflows is not possible based on the present data.

\section{Discussion}
The L988 region contains significant outflow activity as revealed by our surveys for optical and near-IR shock tracers.  The region contains nearly two dozen independent shock systems (20 Herbig-Haro objects and 3 MHO objects unaffiliated with HH objects), however we can only confidently identify the driving protostar for five outflows (\lkha{}324SE, IRAS~21014+5001, WISE~J210214.15+501013.0, WISE~J210222.70+500308.3, and WISE~210223.85+500306.8).  

Two outflows we have identified (from IRAS~21014+5001 and the IRAS~21007+4951 / WISE~J210222.70+500308.3 / WISE~210223.85+500306.8 shock complex) each exceed 1~pc in length with the IRAS~21014+5001 flow being 4.6~pc long.  Both of these large scale outflows originate from sources which are in the highest extinction regions of the cloud.  We find several Herbig-Haro objects and one \mh{} shock complex near the well studied cluster surrounding \lkha{}324 (IRAS~21023+5002, \citeauthor{Clark86} source $e$).

While much of the outflow activity in the L988 cloud is widely distributed over a region $\sim$30\arcmin{} ($\sim$4-7~pc) in diameter, a group of at least three sources which drive overlapping outflows is clustered around the IRAS~21007+4951 source and its associated reflection nebula, located in the center of the L988 cloud.

A notable feature of this study is that outflow activity is detected primarily at visible wavelengths (\ha{} and \sii{}) while our near-IR images (\mh{}) show relatively little shock activity and most of what is detected in the near-IR is coincident with \ha{} or \sii{} shocks.  This is likely due, at least in part, to the combination of the near-IR data being taken on a smaller telescope (3.8 meter as opposed to 8.2 meter) and under much worse seeing conditions ($\sim$1.5\arcsec{} FWHM for the \mh{} images compared to 0.65-0.70\arcsec{} for the \ha{} and \sii{}).  We suspect that further study of this region at near-IR wavelengths will reveal additional details.

\acknowledgments

We would like to thank an anonymous referee for helpful comments.

JW was supported by the NSF through grants AST-0507784 and AST-0407005.  

JW and BR acknowledge support from  the National Aeronautics and Space Administration through the NASA Astrobiology Institute under Cooperative Agreement No. NNA09DA77A issued through the Office of Space Science.

This work is based in part on data collected at the Subaru telescope, which is operated by the National Astronomical Observatory of Japan (NAOJ).  We are grateful to Nobunari Kashikawa for permission to use his [SII] filter for the SuPrimeCam instrument on the Subaru Telescope.  

This publication makes use of data obtained on the United Kingdom Infrared Telescope (UKIRT) which is operated by the Joint Astronomy Centre on behalf of the Science and Technology Facilities Council of the U.K.

This publication makes use of data products from the 2MASS, which is a joint project of the University of Massachusetts and the Infrared Processing and Analysis Center/California Institute of Technology, funded by the National Aeronautics and Space Administration and the National Science Foundation.  

This publication makes use of data products from the Wide-field Infrared Survey Explorer, which is a joint project of the University of California, Los Angeles, and the Jet Propulsion Laboratory/California Institute of Technology, funded by the National Aeronautics and Space Administration.

This research has made use of the VizieR catalogue access tool, CDS, Strasbourg, France. 

MHO catalogue is hosted by Liverpool John Moores University.

We also thank the University of Hawaii Time Allocation Committee for allocating the nights during which these observations were made. The authors wish to recognize and acknowledge the very significant cultural role and reverence that the summit of Mauna Kea has always had within the indigenous Hawaiian community. We are most fortunate to have the opportunity to conduct observations from this sacred mountain.

{\it Facilities:}  \facility{Subaru (Suprime-Cam)}, \facility{UKIRT (WFCAM)}, \facility{UH 88 inch Telescope (Tek2k)}, \facility{WISE}, \facility{2MASS}

\bibliography{refs}

\clearpage
\begin{deluxetable}{llcc}
\tabletypesize{\small}
\tablewidth{0pt}
\tablecolumns{4}
\tablecaption{Positions of Shocks in the L988 region.}
\tablehead{
\colhead{HH Designation} & \colhead{MHO Designation} & \colhead{RA (J2000.0)} & \colhead{Dec (J2000.0)}
}
\startdata
     HH~1044~\phm{A}  & \nodata & 21:01:08.0  & 50:18:39  \\
     HH~1045~A        & \nodata & 21:01:18.6  & 50:02:27  \\
\phm{HH~1045}~B       & \nodata & 21:01:20.8  & 50:02:06  \\
     HH~1046~\phm{A}  & \nodata & 21:01:56.2  & 50:18:44  \\
     HH~1047~A        & \nodata & 21:02:06.3  & 50:09:45  \\
\phm{HH~1047}~B       & \nodata & 21:02:21.0  & 50:11:40  \\
     HH~1048~A        & \nodata & 21:02:26.7  & 50:02:04  \\
\phm{HH~1048}~B       & \nodata & 21:02:28.7  & 50:02:05  \\
\phm{HH~1048}~C       & \nodata & 21:02:30.6  & 50:01:54  \\
\phm{HH~1048}~D       & \nodata & 21:02:31.8  & 50:02:11  \\
\phm{HH~1048}~E       & \nodata & 21:02:33.8  & 50:02:17  \\
     HH~1049~\phm{A}  & \nodata & 21:02:32.9  & 50:14:50  \\
     HH~1050~A        & \nodata & 21:02:34.8  & 50:03:34  \\
\phm{HH~1050}~B       & MHO~958 & 21:02:33.5  & 50:03:13  \\
\phm{HH~1050}~C       & MHO~958 & 21:02:32.1  & 50:03:11  \\
\phm{HH~1050}~D       & MHO~958 & 21:02:28.4  & 50:03:17  \\
\phm{HH~1050}~E       & MHO~958 & 21:02:27.0  & 50:03:03  \\
\phm{HH~1050}~F       & MHO~958 & 21:02:22.1  & 50:03:06  \\
\phm{HH~1050}~G       & \nodata & 21:02:18.6  & 50:03:12  \\
\phm{HH~1050}~H       & \nodata & 21:02:15.1  & 50:02:57  \\
\phm{HH~1050}~I       & \nodata & 21:02:13.5  & 50:03:00  \\
\phm{HH~1050}~J       & MHO~958 & 21:02:12.4  & 50:02:35  \\
\phm{HH~1050}~K       & \nodata & 21:02:11.3  & 50:02:08  \\
\phm{HH~1050}~L       & \nodata & 21:02:09.5  & 50:02:55  \\
\phm{HH~1050}~M       & \nodata & 21:02:08.2  & 50:03:32  \\
\phm{HH~1050}~N       & \nodata & 21:02:07.8  & 50:02:22  \\
\phm{HH~1050}~O       & \nodata & 21:02:07.8  & 50:02:41  \\
\phm{HH~1050}~P       & \nodata & 21:02:04.7  & 50:02:41  \\
     HH~1051~A        & \nodata & 21:02:24.3  & 50:03:09  \\
\phm{HH~1051}~B       & \nodata & 21:02:25.2  & 50:03:15  \\
\phm{HH~1051}~C       & \nodata & 21:02:30.9  & 50:04:01  \\
\phm{HH~1051}~D       & \nodata & 21:02:34.4  & 50:04:08  \\
     HH~1052~A        & MHO~956 & 21:02:38.8  & 50:13:09  \\
\phm{HH~1052}~B       & MHO~956 & 21:02:37.2  & 50:12:57  \\
     HH~1053~\phm{A}  & \nodata & 21:02:50.5  & 50:13:50  \\
     HH~1054~\phm{A}  & MHO~957 & 21:02:50.5  & 50:09:49  \\
     HH~1055~A        & \nodata & 21:03:11.8  & 50:06:19  \\
\phm{HH~1055}~B       & \nodata & 21:03:14.6  & 50:06:01  \\
\phm{HH~1055}~C       & \nodata & 21:03:13.4  & 50:05:55  \\
\phm{HH~1055}~D       & \nodata & 21:03:12.3  & 50:05:37  \\
\phm{HH~1055}~E       & \nodata & 21:03:07.9  & 50:05:30  \\
     HH~1056~A        & \nodata & 21:03:14.9  & 50:15:46  \\
\phm{HH~1056}~B       & \nodata & 21:03:14.9  & 50:14:49  \\
     HH~1057~\phm{A}  & \nodata & 21:03:31.8  & 50:14:46  \\
     HH~1058~\phm{A}  & \nodata & 21:03:38.8  & 50:15:57  \\
     HH~1059~A        & \nodata & 21:03:51.4  & 50:15:57  \\
\phm{HH~1059}~B       & \nodata & 21:03:50.8  & 50:15:53  \\
     HH~1060~\phm{A}  & \nodata & 21:03:52.5  & 50:10:07  \\
     HH~1061~A        & \nodata & 21:04:00.7  & 50:13:55  \\
\phm{HH~1061}~B       & \nodata & 21:04:05.0  & 50:13:18  \\
\phm{HH~1061}~C       & \nodata & 21:04:06.7  & 50:13:10  \\
\phm{HH~1061}~D       & \nodata & 21:04:08.5  & 50:12:52  \\
     HH~1062~A        & \nodata & 21:04:24.9  & 50:13:57  \\
\phm{HH~1062}~B       & \nodata & 21:04:34.5  & 50:13:28  \\
\phm{HH~1062}~C       & \nodata & 21:04:34.9  & 50:13:15  \\
     HH~1063~A        & \nodata & 21:04:27.4  & 50:08:12  \\
\phm{HH~1063}~B       & \nodata & 21:04:27.8  & 50:07:51  \\
\nodata               & MHO~955 & 21:03:14.0  & 50:12:49  \\
\nodata               & MHO~955 & 21:03:14.6  & 50:12:59  \\
\nodata               & MHO~959 & 21:03:48.2  & 50:12:27  \\
\nodata               & MHO~954 & 21:03:51.3  & 50:14:50  \\
\nodata               & MHO~954 & 21:03:49.8  & 50:14:36  \\
\nodata               & MHO~954 & 21:03:52.4  & 50:14:48  \\
\nodata               & MHO~954 & 21:03:50.2  & 50:15:03  \\
\nodata               & MHO~954 & 21:03:49.9  & 50:15:11  \\
\nodata               & MHO~954 & 21:03:50.5  & 50:15:00  \\
\enddata
\label{TableShocks}
\end{deluxetable}

\begin{deluxetable}{lccccccccccc}
\tablecolumns{12}
\tabletypesize{\scriptsize}       
\rotate                                                                                                                                   
\tablecaption{Photometry of Sources in L988.}
\tablehead{
\colhead{Mission} & \colhead{2MASS} & \colhead{2MASS} &\colhead{2MASS} & \colhead{WISE}     & \colhead{WISE}     & \colhead{WISE}    & \colhead{WISE}    & \colhead{IRAS}    & \colhead{IRAS}    & \colhead{IRAS}     & \colhead{IRAS} \\
\colhead{Band}    & \colhead{J}     & \colhead{H}     &\colhead{K}     & \colhead{3.4\um{}} & \colhead{4.6\um{}} & \colhead{12\um{}} & \colhead{22\um{}} & \colhead{12\um{}} & \colhead{25\um{}} & \colhead{60\um{}}  & \colhead{100\um{}} \\
\colhead{(Units)} & \colhead{(mag)} & \colhead{(mag)} &\colhead{(mag)} & \colhead{(mag)}    & \colhead{(mag)}    & \colhead{(mag)}   & \colhead{(mag)}   & \colhead{(Jy)}    & \colhead{(Jy)}    & \colhead{(Jy)}     & \colhead{(Jy)} \\
}
\startdata
LkH$\alpha$~324SE\tablenotemark{a}        & 10.56     &  8.27     &  6.46     & 3.94       & 2.20       & 0.33       & -1.337     & 23.9      & 33.7       & 87       & 199      \\
-                                         & $\pm$0.02 & $\pm$0.02 & $\pm$0.02 & $\pm$0.10  & $\pm$0.09  & $\pm$0.02  & $\pm$0.008 & $\pm$1.2  & $\pm$1.0   & $\pm$11  & $\pm$44  \\
IRAS 21028+5001\tablenotemark{b}          & 10.26     &  9.16     &  8.31     & 7.10       & 6.46       & 4.33       & 2.43       & 0.81      &  1.1       & $<$2.76  & $<$199   \\
-                                         & $\pm$0.02 & $\pm$0.02 & $\pm$0.02 & $\pm$0.03  & $\pm$0.02  & $\pm$0.01  & $\pm$0.01  & $\pm$0.05 & $\pm$0.08  & \nodata  & \nodata  \\
WISE J210339.46+501552.9\tablenotemark{c} & 12.76     & 11.89     & 11.45     & 11.07      & 10.61      & 8.78       & 7.1        & \nodata   & \nodata    & \nodata  & \nodata  \\
-                                         & $\pm$0.02 & $\pm$0.02 & $\pm$0.02 & $\pm$0.02  & $\pm$0.02  & $\pm$0.04  & $\pm$0.2   & \nodata   & \nodata    & \nodata  & \nodata  \\
IRAS 21014+5001\tablenotemark{d}          & 10.10     & 9.95      & 9.86      & 9.57       & 9.32       & 4.34       & 2.08       & 1.12      & 1.71       & 11.8     & 39.8     \\
-                                         & $\pm$0.02 & $\pm$0.02 & $\pm$0.02 & $\pm$0.02  & $\pm$0.02  & $\pm$0.01  & $\pm$0.02  & $\pm$0.06 & $\pm$0.10  & $\pm$1.3 & $\pm$3.2 \\
WISE J210240.13+501236.5                  & \nodata   & \nodata   & \nodata   & 12.48      & 10.69      & 7.78       & 2.38       & \nodata   & \nodata    & \nodata  & \nodata  \\
-                                         & \nodata   & \nodata   & \nodata   & $\pm$0.03  & $\pm$0.02  & $\pm$0.03  & $\pm$0.02  & \nodata   & \nodata    & \nodata  & \nodata  \\
IRAS 21010+5000\tablenotemark{e}          & $<$16.98  & 14.42     & 12.17     & 11.39      & 9.11       & 6.09       & 2.53       & $<$0.36   & 0.81       & $<$11.8  & $<$47.0  \\
-                                         & \nodata   & $\pm$0.05 & $\pm$0.03 & $\pm$0.02  & $\pm$0.02  & $\pm$0.02  & $\pm$0.02  & \nodata   & $\pm$0.08  & \nodata  & \nodata  \\
WISE J210214.15+501013.0\tablenotemark{f} & 14.54     & 13.63     & 13.05     & 11.69      & 10.71      & 8.62       & 7.29       & \nodata   & \nodata    & \nodata  & \nodata  \\
-                                         & $\pm$0.04 & $\pm$0.04 & $\pm$0.05 & $\pm$0.04  & $\pm$0.03  & $\pm$0.04  & $\pm$0.15  & \nodata   & \nodata    & \nodata  & \nodata  \\
2MASS 21024889+5010351                    & 16.32     & 13.42     & 13.94     & \nodata    & \nodata    & \nodata    & \nodata    & \nodata   & \nodata    & \nodata  & \nodata  \\
-                                         & $\pm$0.14 & $\pm$null & $\pm$0.08 & \nodata    & \nodata    & \nodata    & \nodata    & \nodata   & \nodata    & \nodata  & \nodata  \\
WISE J210249.68+501041.9                  & \nodata   & \nodata   & \nodata   & 13.55      & 11.06      & 8.44       & 4.38       & \nodata   & \nodata    & \nodata  & \nodata  \\
-                                         & \nodata   & \nodata   & \nodata   & $\pm$0.03  & $\pm$0.02  & $\pm$0.03  & $\pm$0.03  & \nodata   & \nodata    & \nodata  & \nodata  \\
WISE J210222.70+500308.3\tablenotemark{g} & $<$16.37  & $<$14.82  & 13.28     & 10.72      & 7.16       & 4.23       & 0.90       & \nodata   & \nodata    & \nodata  & \nodata  \\
-                                         & \nodata   & \nodata   & $\pm$0.05 & $\pm$0.03  & $\pm$0.02  & $\pm$0.02  & $\pm$0.01  & \nodata   & \nodata    & \nodata  & \nodata  \\
WISE J210223.85+500306.8\tablenotemark{h} & 12.98     & 11.53     & 10.54     & 9.21       & 7.95       & 5.17       & 2.72       & \nodata   & \nodata    & \nodata  & \nodata  \\
-                                         & $\pm$0.02 & $\pm$0.02 & $\pm$0.02 & $\pm$0.02  & $\pm$0.02  & $\pm$0.02  & $\pm$0.02  & \nodata   & \nodata    & \nodata  & \nodata  \\
IRAS 21007+4951\tablenotemark{i}          & \nodata   & \nodata   & \nodata   & \nodata    & \nodata    & \nodata    & \nodata    & 0.91      & 5.07       & 21.8     & 31.6     \\
-                                         & \nodata   & \nodata   & \nodata   & \nodata    & \nodata    & \nodata    & \nodata    & $\pm$0.12 & $\pm$0.35  & $\pm$2.0 & $\pm$2.5 \\
WISE J210226.45+500203.4                  & 14.81     & 13.05     & 11.66     & 9.19       & 8.06       & 5.60       & 2.71       & \nodata   & \nodata    & \nodata  & \nodata  \\
-                                         & $\pm$0.05 & $\pm$0.03 & $\pm$0.02 & $\pm$0.006 & $\pm$0.02  & $\pm$0.014 & $\pm$0.01  & \nodata   & \nodata    & \nodata  & \nodata  \\
\tablenotetext{a}{Clark source $e$, IRAS 21023+5002, WISE J210358.18+501439.9, see Fig. \ref{RobitalleModels}a}
\tablenotetext{b}{WISE J210428.01+501348.5, see Fig. \ref{RobitalleModels}b}
\tablenotetext{c}{see Fig. \ref{RobitalleModels}c}
\tablenotetext{d}{Clark source $c$, WISE J210303.24+501312.4, see Fig. \ref{RobitalleModels}d}
\tablenotetext{e}{WISE J210242.41+501227.8, see Fig. \ref{RobitalleModels}e}
\tablenotetext{f}{Clark source $f$, see Fig. \ref{RobitalleModels}f}
\tablenotetext{g}{Western Star, see Fig. \ref{RobitalleModels}g}
\tablenotetext{h}{Eastern Star, see Fig. \ref{RobitalleModels}h}
\tablenotetext{i}{Clark source $a$.  This IRAS source may be a blend of the two sources listed above it (WISE J210222.70+500308.3 \& WISE J210223.85+500306.8).}
\enddata
\label{TableSources}
\end{deluxetable}

\end{document}